\documentclass[]{raa}            
\usepackage{graphicx,times}
\usepackage{natbib}

\providecommand{\bm}[1]{\mbox{\boldmath$#1$\unboldmath}}

\begin{document}

\title{A Quantity Characterising Variation of Observed Magnetic Twist of Solar Active Regions
} \volnopage{ {\bf ????} Vol.\ {\bf ?} No. {\bf XX}, 000--000}
   \setcounter{page}{1}
   \author{Y. Gao
      \inst{1}
   }
   \institute{Key Laboratory of Solar Activity, National Astronomical
Observatories, National Astronomical Observatories, Chinese Academy
of Sciences,
             Beijing 100012, China; {\it gy@bao.ac.cn}\\
\vs \no
   {\small Received [year] [month] [day]; accepted [year] [month] [day] }
} \abstract{An alternative parameter $R_{J_z}$ is introduced as the ratio of one of two kinds
of opposite-sign current to the total current and investigate the
relationship between the quantity and the hemispheric sign rule of
helicity (HSR) that is established by a series of previous statistical
studies. The classification of current in each hemisphere is according
to the following rule: If the product of the current and the corresponding
longitudinal field component contributes a consistent sign with reference to the HSR,
it is called ``HSR-compliant" current, or else it is called ``HSR-noncompliant" current.
Firstly, the consistence between the butterfly diagram of the $R_{J_z}$
and the current helicity was obtained in a statistical study. Active
regions with $R_{J_z}$ smaller than 0.5 tend to obey the HSR whereas
those with $R_{J_z}$ greater than 0.5 tend to disobey the HSR. The
``HSR-compliant'' current systems have 60\% probability of realization
compared to 40\% of ``HSR-noncompliant'' current systems. Overall, the HSR
is violated for active regions in which the ``HSR-noncompliant" current is
greater than the ``HSR-compliant" current.
Secondly, the $R_{J_z}$ parameter was subsequently used to study
 the evolution of current systems in the case analyses of
flare-productive active regions NOAA AR 11158 and 11283. It is found
that there were ``$R_{J_z}$-quasi-stationary" phase that is relatively
flare quiescent and ``$R_{J_z}$-dynamic" phase that
is covered by the occurrence of large flares.}

 \keywords{sun: activity --- sun: flares --- sun: magnetic field ---sun:evolution --- sun: sunspots}

   \authorrunning{Y. Gao}            
   \titlerunning{A Quantity Characterising Magnetic Twist}  
   \maketitle


%
%
\section{Introduction}

The chirality of active region magnetic fields has been
studied in terms of the current helicity or the linear force-free
field $\alpha$. In the northern (southern) solar hemisphere, there is statistically negative (positive) sign preference of helicity quantities; this trend is called the hemispheric helicity sign rule
\citep[HSR hereafter in this paper; { see}][{ for results based on
the data before solar cycle
24}]{Seehafer90,Pevtsov94,Pevtsov95,Abramenko96,BaoZ98,HS04,Tiwari09,Zhang10}.
To be related closely, some case analyses showed that there were opposite electric current systems in several active regions \citep{WangT94, Leka96, WangT99, Wheatland00}. This implies that both left and right handedness of field coexist in active regions, as have been also supported by the observation of \citet{Su09}, an $\alpha$ map may contain mixed signs in sunspots. In addition, the statistical study of helicity sign was carried out in the solar minimum \citep[e.g.,][]{HZ12}. Recently, the HSR has been further confirmed with the studies on helicity injection from emerging active regions \citep{Yang09, ZhangY13} and \textit{Solar Dynamics Observatory }{\it (SDO)}/Helioseismic and Magnetic Imager (HMI) observation by \citet{Liu14}. These studies have shown that the HSR has large dispersion; the ratio of preferred helicity sign is about 60\%.

On the other hand, some results showed that the HSR might not hold
throughout the solar cycle \citep{Baoetal00, HS05, Zhang10}. This
observational characteristic was obviously important for the
theoretical formulation of dynamo models \citep{Choudhuri04,
Pipin13}. It has been further confirmed that there were net current above both polarities of magnetic field in several active regions \citep{Gao13a}.

The study has indicated that the HSR can also
be investigated in term of electric current distributions in active
regions. Corresponding to the locus of sign reversal of helicity
found in the helicity butterfly diagram by \citet{Zhang10}, there is
also reversal of sign of net electric currents in the butterfly
diagram \citep{Gao13a}. On the other hand, the contribution of
opposite-sign helical fields to the filament eruptions has been
observed by \citep{LiuK04, Shenetal15}. \citet{BY16} found that the reversal direction of a sunspot¡¯s rotation during an X1.6 flare. As a result, we speculate that the currents which do not conform to the HSR have its own significance and it is important to investigate their properties in broader viewpoint of observation.

In this paper a new quantity $R_{J_z}$ is introduced; in a given active region it quantifies the ratio of the currents that do not obey the HSR to the total currents. First, the solar-cycle evolution of $R_{J_z}$ will be studied and used to interpret the distribution of the helicity in the butterfly diagram in Section 2 from the viewpoint of dynamic evolution of two kinds of electric current. In Section 3, using $R_{J_z}$, it was further investigated of the evolution of current systems above opposite polarities in two active regions that produced large flares.

\section{Data Observed at Huairou Solar Observing Station}
\subsection{Definition of Parameters}

Our starting point is the well-known definition of the electric
current:

\begin{equation}
J_z= (1/\mu_{0}) (\partial B_{y}/\partial x-\partial B_{x}/\partial
y)
\end{equation}
where $\mu_{0}=4 \pi \times 10^{-3}$ G m A$^{-1}$. We define the
ratio of the currents:

\begin{equation}
R_{J_z}^{\pm}=-
\frac{\sum_{i}J_{z_i}}{\sum_{j}J_{z_j}-\sum_{i}J_{z_i}}
\label{eq:RJz}
\end{equation}

where ``$i$" (``$j$") denote the pixels that have sign of the
current helicity $H_c = B_{z} \cdot J_{z}$ that is inconsistent
(consistent) with the sign of the current helicity according to the
HSR. If $\theta$ is the latitude of a region under consideration,
then ``$i$" pixels are in regions with $\theta B_z J_z > 0$ where
``$j$" pixels are $\theta B_z J_z < 0$. According to the HSR, ``$i$"
represents the ``HSR-noncompliant" current system and ``$j$" represents the
``HSR-compliant" current system. The superscript $\pm$ means that $R_{J_z}$
is computed in regions of positive ($+$) or negative ($-$) magnetic
polarity, respectively. In either positive or negative polarity
regions, $J_{z_i}$ and $J_{z_j}$ have opposite signs, and $R_{J_z}$
is always positive and between 0 and 1. The parameter $R_{J_z}$
represents the fraction of ``HSR-noncompliant" currents normalized by the total currents. Likewise, $1-R_{J_z}$ represents
the fraction of ``HSR-compliant" currents and the difference between the two
is $1-2R_{J_z}$. The latter quantity might be compared with the
helicity imbalance parameter $\rho_h$ introduced by \citet{BaoZ98}
if we consider the current helicity instead of the current itself.

If $R_{J_z}$ is defined as a fraction of pixels with consistent sign of $H_c$, it will be equivalent to $1-R_{J_z}$. In my case of definition, the value of $R_{J_z}$ in each HSR-compliant region tends to be less than 50\%, and that in each HSR-noncompliant region tends to be greater than 50\%. The distribution of electric current with the longitudinal magnetic field affects the ultimate determination of sign of $\langle H_c \rangle$ in an active region, as shown in Figures 2 (b) and (c) later. This definition is concerned with the physical scenario of observed opposite-sign current in the magnetogram. How to understand the current which does not conform to the HSR is still an open question. In this paper, the significance of the $H_c$-noncompliant current in the observation is in agreement with the fact that the HSR rule has a dispersion presently restricted only to 60\% of ARs. That is to say, the variation of $H_c$-compliant and $H_c$-noncompliant current system is probably more intrinsic property of evolution of twist in solar active regions.

To further declare the quantity of $R_{J_z}$ and understand its relation with the current helicity, it was used of the current helicity butterfly diagram with HSOS vector magnetograms over two solar magnetic cycles \citep{Zhang10}. However, when we study the evolution of $R_{J_z}$ in individual active region, the high-cadence vector magnetograms observed by the spatial instrument as \emph{SDO}/HMI are necessary.

\subsection{Connection of $R_{J_z}$ to Previous Statistical Observation}
\label{sect:Obs} The vector magnetograms used here have been obtained
with the Solar Magnetic Field Telescope (SMFT) of Huairou Solar
Observing Station, National Astronomical Observatories of China.
The basic information of the instrument can be referred to \citet{BaoZ98}.
Following \citet{Zhang10}, in computations of the
current helicity, the pixels with signal that exceeds the
noise levels ($|B_{z}|>$ 20 G and $B_{t}>$100 G) were used.

\label{sect:data}

\citet{Gao13a} has shown that the net electric currents follow a
butterfly-diagram-like evolution over the solar cycle. The analysis
of 6629 vector magnetograms observed at Huairou Solar Observing
Station from 1988 to 2005 has revealed that $R_{J_z}$ also shows a
butterfly diagram (see Figures 1 and 2). Detailed information about
the data used for the production of Figures 1 is given in
\citet{Zhang10}. In Figure 1, the color background represents the
values of $R_{J_z}^{+}$ (Figure1a) and $R_{J_z}^{-}$ (Figure 1b).
These values have been averaged over the same intervals in time and
latitude as the current helicity. The overplotted filled or open
circles are the averaged current helicity $\langle H_c \rangle$ from
the same data sample. The sizes of open and filled circles in Figure
1 are different from \citet{Zhang10} because a different
way of display was adopted. The size of open or filled
circles is scaled by using the ratio of the value in each bin to the maximum
of absolute value that can be seen from the label. Meanwhile, the square
root value for each bin was adopt so that the sizes of open
and filled circles can be visually comparable. From the figure, we can see:

\begin{enumerate}

  \item $R_{J_z}^{+}$ in Figure 1a and $R_{J_z}^{-}$ in Figure 1b show
  similar patterns. The correlation between $R_{J_z}^{+}$ and
  $R_{J_z}^{-}$ is shown in Figure 3(a) with a linear correlation
  coefficient of 0.52. This value is highly significant since the
  level of 99\% significance for the two-tailed test for 100 degrees
  of freedom is 0.254, while the total number of the studied active
  regions is 983. Both average values of $R_{J_z}$ on the two
  polarities are around 0.48.

 \item The color of the background for $R_{J_z}$ in Figure 1 tends to be
 green $R_{J_z}<0.5$ in accordance with the overplotted filled circles
 in the north hemisphere and open circles in the south hemisphere.
 On the contrary, the color of the background tends to
 blue $R_{J_z}>0.5$ in accordance with
 the sizes of overplotted open circles in the north hemisphere and
 filled circles in the south hemisphere. That is to
 say, regions with the $R_{J_z}$ less (greater) than 0.5 tends to obey
 (disobey) the HSR.

 \item The sign reversal of helicity tends to occur where the fraction
  of ``HSR-noncompliant" current systems is \emph{greater than 0.5},
  manifested in blue colors in Figure 1. The two boxes in black in Figure 1 show a typical
  example. For more quantitative analyses, we use the sign function of
  latitude, and consider the quantity sign $(\theta) \cdot \langle H_c \rangle$.
  If the helicity obeys/disobeys the HSR, sign $(\theta) \cdot \langle H_c \rangle$
  will be negative/positive. Thus we expect that the $R_{J_z}$ is
positively correlated with sign $(\theta) \cdot \langle H_c
  \rangle$. Figures 2(b) and 2(c) show that this is really the case.
  The well correlation between the magnitude of the $R_{J_z}$ and $\langle H_c
  \rangle$ indicates the sign of $\langle H_c
  \rangle$ reflects the chirality of field than the field strength.

  \item We computed the percentage of active region numbers in each
  quadrant of Figures 2(b) and 2(c), and found that they are 30.6\%
   (first quadrant), 10.4\% (second quadrant), 49.1\% (third
   quadrant) and 9.8\% (fourth quadrant), respectively, in Figure 2(b),
   The active regions in the second and third quadrants obey
   the HSR, which are in total 59.5\%. Figure 2(c) shows similar
   number; the percentages of active regions in four quadrants are
   29.1\%, 11.4\%, 48.1\% and 11.4\%, respectively, and again 59.5\%
   of active regions follow the HSR. The first quadrant shows that
   the $R_{J_z}$ and $\langle H_c \rangle$ both disobey the HSR, while the third quadrant shows that the $R_{J_z}$ and $\langle H_c \rangle$ both obey the HSR. The percentages are around 30\% for the first quadrant and 50\% for the third quadrant. Particularly, the second quadrant shows that the $R_{J_z}$ disobeys the HSR but the $\langle H_c \rangle$ obeys that the HSR, on the contrary, the fourth quadrant shows that the $R_{J_z}$ obeys the HSR but the $\langle H_c \rangle$ disobeys the HSR. The percentages in the second and fourth quadrant are around 10\%. These two parts imply the distribution of electric current with the longitudinal magnetic field affects the ultimate determination of sign of $\langle H_c \rangle$ in an active region.

   \item Statistically, when $R_{J_z}$ is less (greater) than 0.5,
   the active region obeys (disobeys) the HSR. As an active region
   evolves the current systems evolve as well, and the ``HSR-compliant" current
   systems may become smaller than the ``HSR-noncompliant" one so that the active
   region disobeys the HSR, and vice versa. However, over the whole
   solar activity cycle, in 60\% of the active regions the ``HSR-compliant"
   current system is greater than the ``HSR-noncompliant" one, hence
   accounting for the HSR. So from the viewpoint of two current system
   of opposite sign coexisting in the same magnetic polarity,
   it can also account for the hemispheric sign rule of helicity.

\end{enumerate}

\section{Case Study Using the Data from {\it SDO}/HMI}
\subsection{Detailed Analyses on Two Active Regions}

The new generation vector magnetograph {\it SDO}/HMI provides more stable
time-series than ground-based ones and allow us to study whether there is
the relative variation of current system of solar magnetic field with time.
If the $R_{J_z}$ reveals
the variation of real chirality of magnetic field, it is expected to
see the evolution of two kinds of current systems. To this end,
we studied two flare-productive active regions using vector
magnetograms obtained with {\it SDO}/HMI firstly.

HMI's basic information can be referred to \citet{Schou12}. It contains a full disk ($4096 \times 4096$) filtergraph with a pixel resolution of 0.5
arcsec.The working spectral line is Fe {\sc i} 617.3 nm line through
a 76 m\AA\ passband filter at six wavelength positions across the
line. The processing of vector magnetic field by using the Very Fast Inversion of the Stokes Vector algorithm based on the Milne-Eddington atmospheric model can be referred in Hoeksema et al. (2014). The 180$^{\circ}$ ambiguity of horizontal field was resolved with the minimum energy method (Leka et al. 2009).

The first region studied here is NOAA AR~11158; it was a
well observed rapidly developing active region and widely studied from different viewpoints \citep{Sun12,Jetal12,Nindos12,Song13,Vemareddy15}.
The other region studied here is NOAA~AR~11283;  The detailed information on the flares
referred to this study was same as Table 1 in \citet{Gao14}.

\subsection{Electric Currents in NOAA 11158 and 11283}

Figure 3(b) shows the total electric currents in NOAA
11158. This region was located in the southern hemisphere.
The net currents can be measured by the difference between
the curves of the ``HSR-compliant" and ``HSR-noncompliant"
current in Figure 3(b). The
total net current $\sum(J_{z_j}^{+}+J_{z_i}^{+})$ above the positive
(negative) is positive (negative). Therefore, AR11158 obeyed the
HSR.

Figure 4(b) shows the evolution of total electric currents
in NOAA 11283. This region was located in the northern hemisphere.
If it obeys the HSR, it should have negative helicity. From Figure
4(b), we can see that the net currents were positive in the negative
polarity and negative in the positive polarity, i.e. negative
helicity, before September 5. The signs of the currents changed
after September 5, leading to positive helicity, against the HSR.
This is consistent with the results obtained by \citet{GZZ12}.

The opposite signs of net electric current above opposite
magnetic polarity in Figure 3(c) and 4(c) agree with the results
 obtained by \citep{Gao13a}. Furthermore,
they indicate significant changes in the current systems during the
evolution of the regions; the net currents above regions of opposite
magnetic polarities show variations in almost precisely the opposite
sense, indicating a closure of current systems flowing between the
two polarities. The unit of electric current density was used for $\Sigma J_z$ and $\langle J_z \rangle$, that stand for the integrated and averaged magnitude of electric current density over all of selected pixels. The main difference with the unit of electric current applied in some other analyses (e.g., Vemareddy, Venkatakrishnan and Karthikreddy, 2015; Vemareddy, Cheng and Ravindra, 2016) is the factor of the area of each pixel, 2.54 $\times$ $10^{11}m^2$ for these two sets of HMI vector magnetograms with spatial resolution of 0.504$^{''}$ per pixel.

\subsection{$R_{J_z}$ in NOAA 11158 and 11283}

Figure 3(d) shows the evolution of $R_{J_z}$ in NOAA AR 11158.
 The corresponding error is estimated with the method of Monte Carlo
simulation. By adding a random noise that is less than that
recorded error of each measured vector magnetic field, then we
repeat the computation many times and get the final average
value and the corresponding standard deviation. During the time
interval we studied, the value of $R_{J_z}^{+}$
(red) increased from a minimum of 0.431 at 09:24 UT on February 13
{(\it some 8.06 h before the first M6.6 flare)} to a maximum of
0.490 at 18:00 UT on February 14 and then decreased to 0.449 at
15:36 UT on February 15 {\it (some 14 h after the X2.2-class
flare)}. The evolution of $R_{J_z}^{-}$ (blue) shows similar
characteristics. The linear correlation coefficient between
$R_{J_z}^{+}$ and $R_{J_z}^{-}$ is 0.85. Such a high correlation
indicates a coherent variation of electric currents on the opposite
polarity regions during the evolution of this region.

Figure 4(d) shows the evolution of the $R_{J_z}$ in NOAA AR 11283.
During the time interval we studied, the value of
$R_{J_z}^{+}$ increased from a global minimum of 0.484 at 21:36 UT
on September 04 (about one day before the first M5.3 flare) to a
maximum of 0.533 at 08:48 UT on September 07 and then decreased to
reach 0.513 at 11:48 UT on September 08 again (about 12 h after the
X1.8 flare). The subsequent evolution of $R_{J_z}$ is not known
because no vector magnetograms were available. The time profiles of
the $R_{J_z}^{+}$ and $R_{J_z}^{-}$ are similar; the linear
correlation coefficient between the two quantities is 0.81.

The above results may indicate that the time variation in $R_{J_z}$
can be used to identify an ``$R_{J_z}$-dynamic" phase. Therefore,
we divide the studied periods into two intervals: ``$R_{J_z}$-quasi-stationary"
and ``$R_{J_z}$-dynamic" phases. The start of the ``$R_{J_z}$-dynamic"
phase is taken at the time when $R_{J_{z}}$ begins to rise
to a global maximum. The end of the ``$R_{J_z}$-dynamic" phase is
taken as the time when $R_{J_{z}}$ returned to a local minimum.
Hence the ``$R_{J_z}$-quasi-stationary" phase is the time outside of the
``$R_{J_z}$-dynamic" phase. In particular, for AR11158 and
AR11283 the ``$R_{J_z}$-dynamic'' phases are the intervals
between the vertical dashed lines in Figures 3(c) and 4(c),
respectively: for AR11158 the ``$R_{J_z}$-dynamic" phase was from
09:24 UT on February 13 to 15:36 UT on February 15 and for AR11283
it was from 21:36 UT on September 04 to 11:48 UT on September 8.
Interestingly, the behavior of $R_{J_z}$ in the ``$R_{J_z}$-dynamic"
phase was similar in the two analyzed active regions.

In order to see where such variations in the
electric currents take place, eight moments were chosen and
marked with arrows in Figures 3(d) and 4(d), and plotted the corresponding
snapshots of the electric currents. The ``P" and ``Q" regions in panels (a-h)
of Figures 5 and 6 show the regions in which prominent variation
of \emph{HSR-noncompliant} current occurred. Here the prominent
variation of current are shown with ``P" and ``Q" regions, but it does not mean the
variation only occurs at these regions but must exist in other places that
are not clearly shown. At least in the region of opposite magnetic polarity,
there is corresponding well correlated variation of current. This can be inferred
from the Figure 3 (d) and 4 (d).

Compared with the evolutional trends of other parameters for AR 11158 (e.g., Song et al. 2013),
such as electric current, current helicity, photospheric free
energy, and angular shear etc., the curve $R_{J_z}$ has a similar rising trend in the former half of ``$R_{J_z}$-dynamic" phase. However, there is an obviously
different descending trend in the latter half of ``$R_{J_z}$-dynamic" phase.
After the ``$R_{J_z}$-dynamic" phase, the $R_{J_z}$ returns to the level as that
before the ``$R_{J_z}$-dynamic" phase.
The decreasing tendency of parameter measuring the free magnetic energy would be expected after the flare, as pointed out by (Wiegelmann, Thalmann and Solanki, 2014). Although there are different trend in some particular situations, these parameters are all important to show the storage and release of magnetic energy in different ways.

When the computation was performed, the pixels where $|B_z| \ge 50 G$
are taken into the final determination of the parameters so that the
uncertainty of horizontal and vertical field outside of the active region would affect little these parameters. To the evolutional curves of AR 11158 and 11283, the error propagation was estimated by Monte Carlo method.
In particular, 30 sets
of parameter were obtained at each moment by adding the random errors to the inputted
field strength. Take $J_z$ for example,
firstly the $J_z$ with account of random error is computed as follows:
$J_{z}= (1/\mu_{0}) \{ \partial [B_{y}+2\times(R_0-0.5)\times\delta B_y]/\partial x-\partial [ B_{x}+2\times(R_1-0.5)\times\delta B_x]/\partial
y\}$,
where $R_0$ and $R_1$ are random number units. $\delta B_{x}$ and
$\delta B_{y}$ are inversion errors of field components provided by HMI.
Then the standard deviation of these 30 sets
of parameter ($\delta_{J_z}$) is taken as the error estimation of this moment. The error at each moment is shown with short bar of the corresponding quantity in (b), (c) and (d) of Figure 3 and 4.

\section{Conclusions and Discussion}
\subsection{Conclusions}
\label{sect:conclusion}

From long-term observation obtained at the Huairou Solar Observing
Station that covered more than 1.5 solar cycles, two current systems
 can be identified: they were named as ``HSR-compliant" and ``HSR-noncompliant"
current systems according to whether their signs conform to the HSR.
It was found that the active regions with $R_{J_z}$ less than 0.5 tend
to obey the HSR while active regions with $R_{J_z}$ greater than 0.5
tend to disobey the ``HSR". It was also found that the ``HSR-compliant" current
system has about 60\% probability of realization greater than the
``HSR-noncompliant" current system. This marginal superiority of the normal
current system may explain the reversal of the helicity sign and big
scatter in the HSR. At present it was uncertain whether the above
picture holds for any given time. From the locus of sign reversal of
helicity in Figures 1, it is inferred that the probability of ``HSR-compliant"
vs. ``HSR-noncompliant" current systems may be different for different active
regions or in different phases of the solar cycle.

Active regions studied here exhibited an ``$R_{J_z}$-dynamic"
phase in the time profiles of their $R_{J_z}$. The ``$R_{J_z}$-dynamic"
phase is the interval characterized by the gradual increase
and then decrease in $R_{J_z}$ on both polarities. In the studied
active regions, large flares occurred during this interval. Eight moments were
chosen that marked in Figure 3 and 4, then the
corresponding snapshot of the electric current was plotted. The green arrows in
panels (a-h) of Figure 5 and 6 indicate the regions of prominent
variation and with the abnormal helicity. This shows a prominent
increase and then decrease in one of the current systems in
the ``$R_{J_z}$-dynamic" phase, violating the HSR for the two
particular active regions in this study. This conjecture still needs
further confirmation by a statistical work with a bigger sample.
However, $R_{J_z}$ could be a sensitive indicator highlighting
peculiar properties of active regions around large flares compared
to their properties in relatively quiescent periods. It was noted that in
the two active regions, their long-term evolution of
$R_{J_z}$ that defined their ``$R_{J_z}$-dynamic'' was similar.
This might indicate that $R_{J_z}$ reflects the underlying physical
process that occurs commonly in current systems of different active
regions around the time of large flares.

The high correlation coefficient between the time profiles of
$R_{J_z}$ that are associated with opposite magnetic polarities was
also found in AR11158 and AR11283. It implies that the current
systems in the opposite polarities evolve coherently, namely the
currents connect the two polarities by flowing basically along the
field lines.

\subsection{Discussion}
\label{sect:discussion}

The $R_{J_z}$ parameter measures the difference in magnitudes of two
opposite current systems which accounts for whether the active
region obeys HSR or not. Besides, it shows large-scale temporal
trends associated with the occurrence of large flares for the two
different active regions studied. This property may be applied to further study
flares in different active regions, though it needs to be stressed
again that a statistical work with more examples is needed to
confirm the universality of the $R_{J_z}$-dynamic phase covering the major flare. Besides, how to quantitatively separate the $R_{J_z}$-quasi-stationary from the $R_{J_z}$-dynamic phase needs further investigation.

It should be pointed out that the dynamic evolution of electric current in the solar active region investigated in the current paper is based on the observation
in space. The time-series of vector magnetogram are obtained with high cadence and snapshots show the evolution of active region independent with the atmosphere around
the earth. Up to now, the \emph{SDO}/HMI instrument provides the unique data for this investigation. Even for this instrument, it was uncertain for how many observed active region we can see the similar variation. This is to present an alternative quantity that possibly reflects the dynamic evolution of two kinds of opposite magnetic twist in individual polarity longitudinal magnetic field by analyzing two well-observed active regions.

\begin{acknowledgements}
I am grateful to the referee for his/her constructive comments that improve
 the manuscript. The work is supported by the National Natural
Science Foundation of China under grants 11103037, 11273034,
11178005, 41174153, 11173033, 11473039 National Basic Research Program of
China under grant 2000078401 and 2006CB806301, and Chinese Academy
of Sciences under grant KJCX2-EW-T07. SDO is a NASA mission, and HMI
project is supported by NASA contract NAS5-02139 to Stanford
University.
\end{acknowledgements}

\clearpage
\begin{figure}
\centering
\includegraphics[angle=90,scale=0.6]{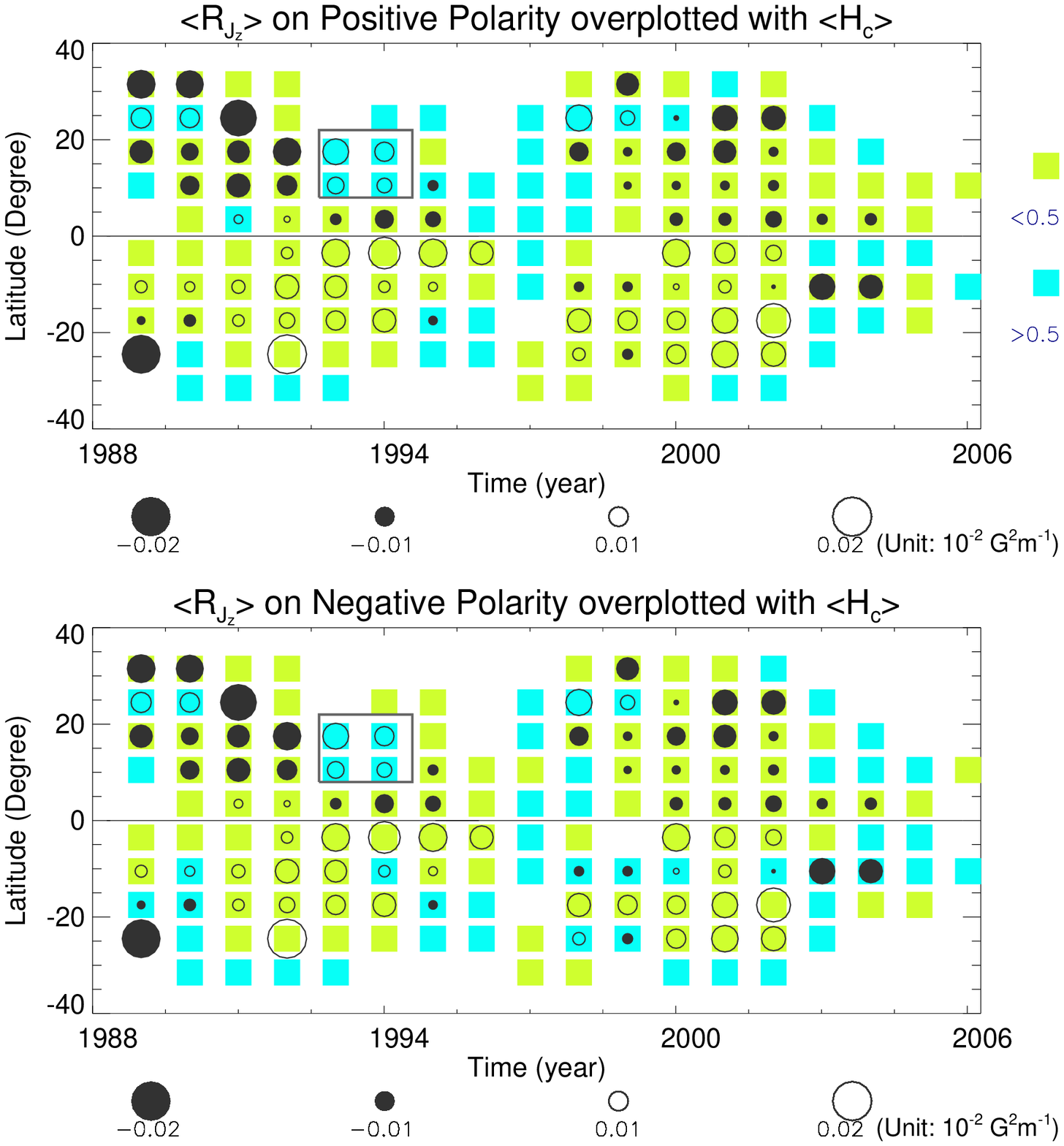}
\caption{(a) Butterfly diagrams of $R_{J_z}$ (colors) associated
with averaged current helicity $\langle H_c \rangle$ (open and
filled circles). The vertical axis gives the latitude and the
horizontal axis gives the time in years. The values of $R_{J_z}$ are
scaled according to the color square that appears to the right of
the panel. The sizes of open/filled circles correspond to the
magnitude of the $\langle H_c \rangle$ according to the scale that
appears under the horizontal axis labels. (b) The same as panel (a) but for $R_{J_z}$
associated with negative magnetic fields. Butterfly diagrams of $\langle H_c \rangle$ was plotted in agreement to that from Zhang et al. (2010) by keeping the same requirement of at least 30 data samples in each latitude-time bin. However, this requirement was released to if only there are 2 data samples in each latitude-time bin for plotting butterfly diagrams of $R_{J_z}$ so that the reversal-sign features on the edges can be shown.} \label{fig1}
\end{figure}

\begin{figure}
\centering
\includegraphics[angle=90,scale=0.6]{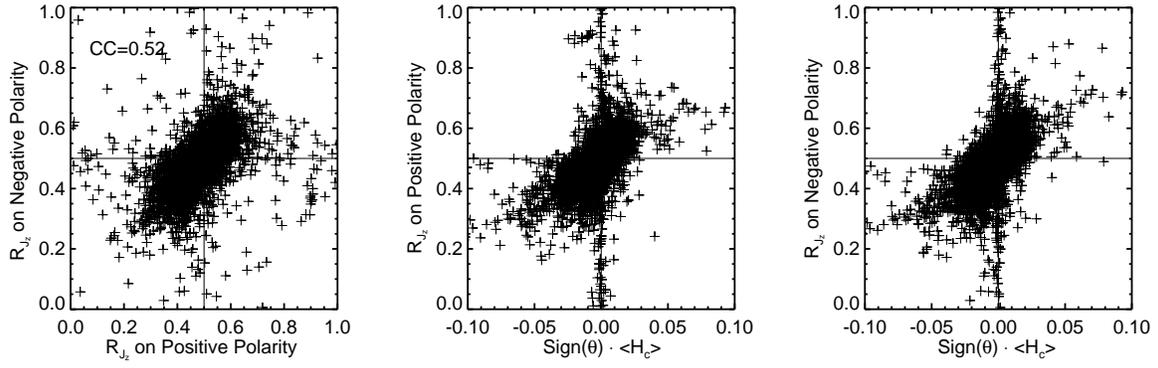}
\caption{(a): Correlation between $R_{J_z}^{+}$ and $R_{J_z}^{-}$.
(b): Correlation between $R_{J_z}^{+}$ and sign ($\theta$) $\cdot$
$\langle H_c \rangle$. (c): Correlation between $R_{J_z}^{-}$ and
sign ($\theta$) $\cdot$ $\langle H_c \rangle$. } \label{fig2}
\end{figure}

\clearpage

\begin{figure}
\centering
\includegraphics[angle=0,scale=0.6]{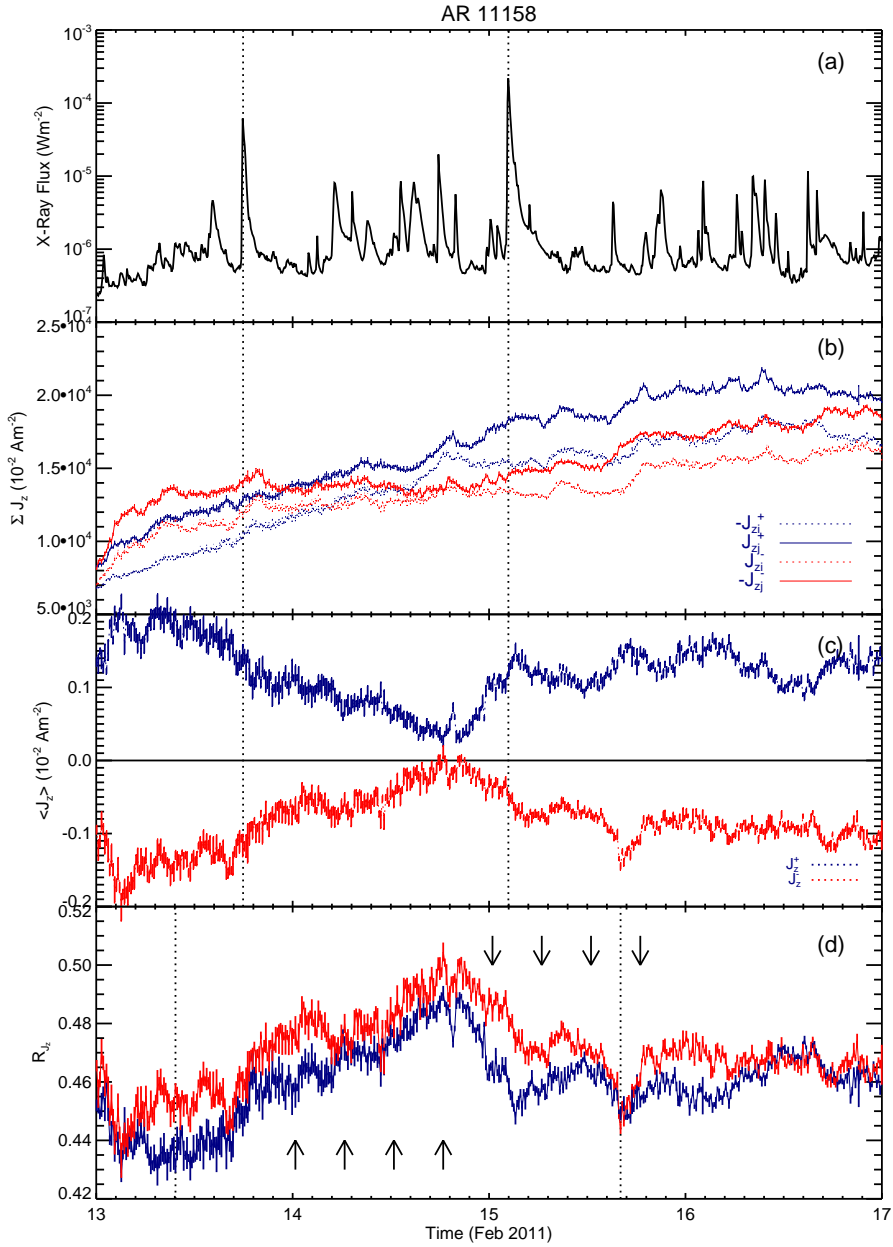}
\caption{(a): GOES
X-ray flux from 2011 February 13 to 17. (b): Evolution of $\sum J_z
$ (red and blue in positive and negative polarities) in NOAA AR
11158. The vertical dashed lines correspond to the times of the
flares discussed in the text. (c): Evolution of $\langle J_z \rangle$ (d): Evolution of $R_{J_z}$. The
interval between the two vertical dashed lines corresponds to the
``$R_{J_z}$-dynamic'' phase discussed in the text.} \label{fig3}
\end{figure}

\clearpage

\begin{figure}
\centering
\includegraphics[angle=0,scale=0.6]{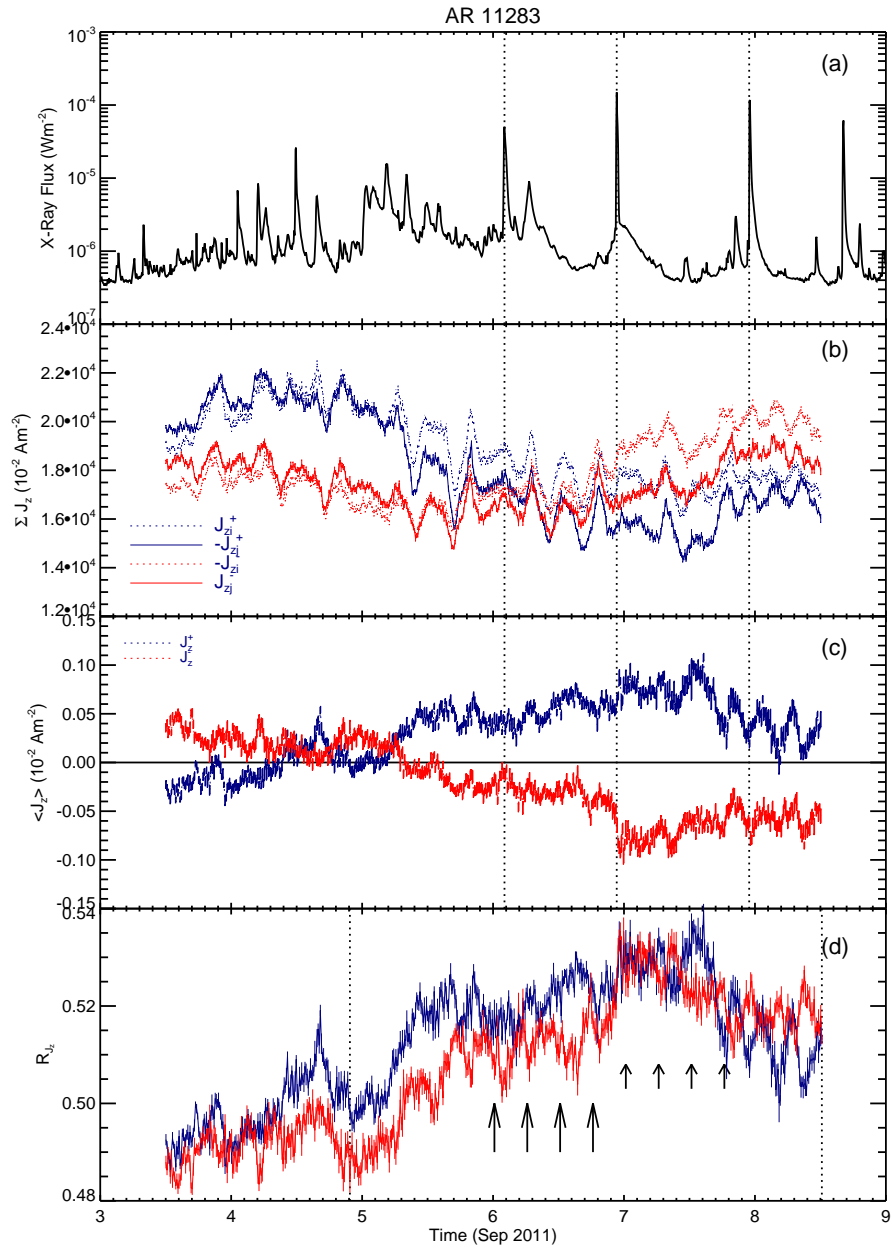}
\caption{Same as
figure 3 but for the GOES X-ray flux from 2011 September 3 to 9 and
 NOAA AR 11283.}
\label{fig4}
\end{figure}

\begin{figure}
\centering
\includegraphics[angle=0,scale=1.9]{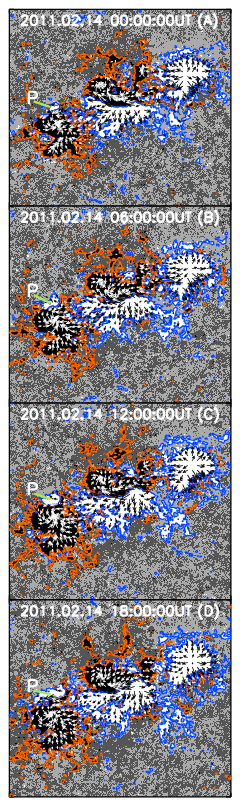}
\includegraphics[angle=0,scale=1.9]{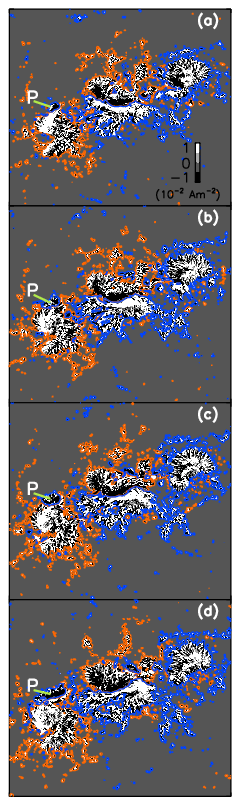}
\end{figure}
\clearpage

\begin{figure}
\centering
\includegraphics[angle=0,scale=1.9]{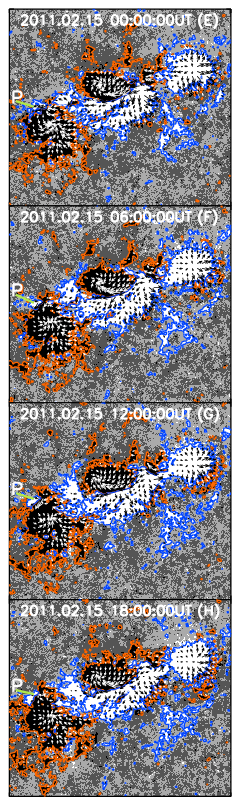}
\includegraphics[angle=0,scale=1.9]{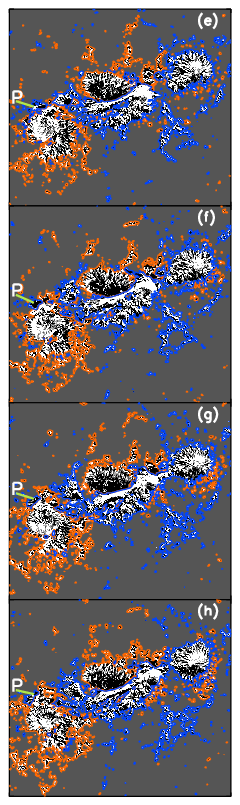}
\caption{The sample of vector magnetograms (A-H) and corresponding maps of $J_z$ (a-h) at eight
selected moments of NOAA AR 11158. The blue (orange) contours show
the levels of $\pm 50G$.}
\label{fig5}
\end{figure}

\begin{figure}
\centering
\includegraphics[angle=0,scale=1.9]{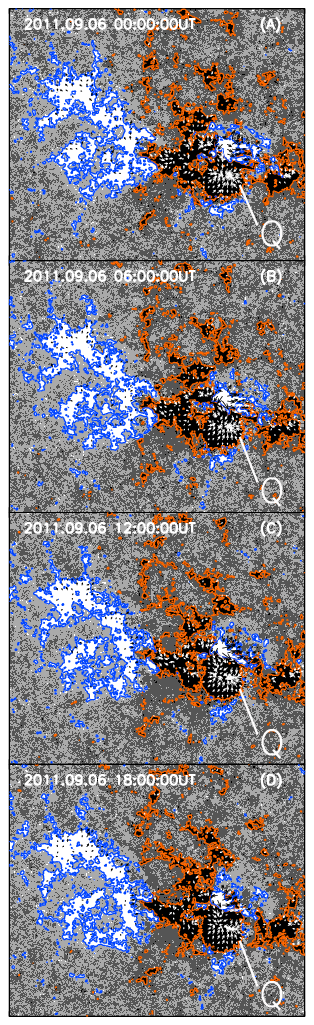}
\includegraphics[angle=0,scale=1.9]{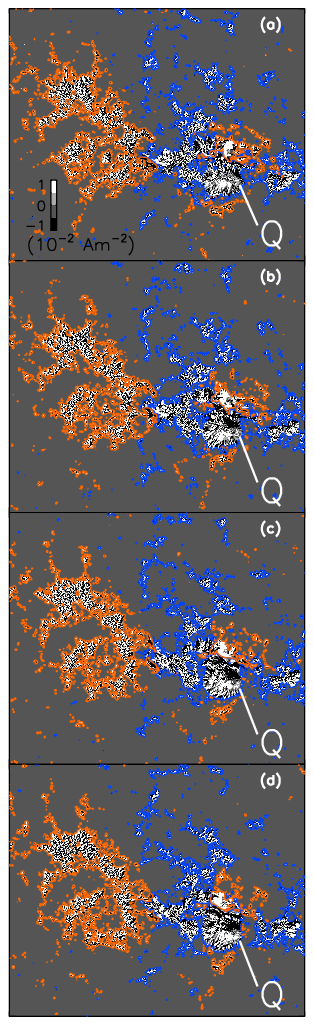}\\
\end{figure}
\clearpage

\begin{figure}
\centering
\includegraphics[angle=0,scale=1.9]{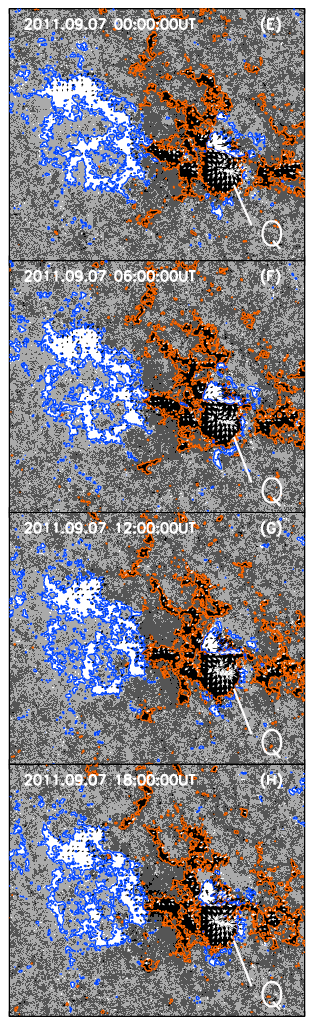}
\includegraphics[angle=0,scale=1.9]{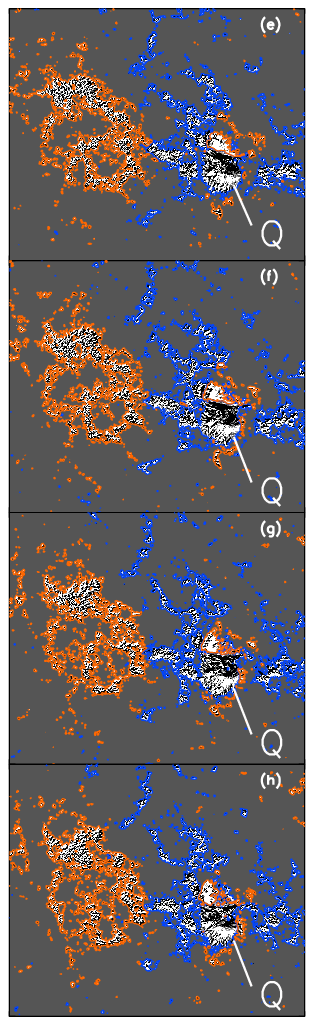}
\caption{The sample of vector magnetograms (A-H) and corresponding maps of $J_z$ (a-h) at eight
selected moments of NOAA AR 11283. The blue (orange) contours show
the levels of $\pm 50G$.}
\label{fig6}
\end{figure}

\label{lastpage}

\end{document}